\documentclass[aps, prl, 10pt, twocolumn, superscriptaddress, notitlepage, 
floatfix, longbibliography]{revtex4-1}
\usepackage{amsmath,amsfonts, amssymb, amsthm, dsfont}
\usepackage[T1]{fontenc}
\usepackage{bm}
\usepackage{graphicx}
\usepackage{verbatim}
\usepackage[bookmarks=true,colorlinks=true,linkcolor=blue, urlcolor=blue, citecolor=blue]{hyperref}
\hypersetup{linktocpage}
\usepackage{xcolor}
\usepackage{bbold}
\usepackage{braket}
\usepackage{environ} 
\usepackage{relsize} 
\usepackage{enumitem}
\usepackage[caption=false]{subfig} 
\usepackage[normalem]{ulem}

\NewEnviron{eqs}{%
\begin{equation}\begin{split}
    \BODY
\end{split}\end{equation}
}

\newcommand{\fTC}{{\text{f\nobreak\hspace{.08em plus .08333em}TC}}}

\def\cC{\mathcal{C}}

\def\cZ{\mathcal{Z}}

\def\cV{\mathcal{V}}
\def\cN{\mathcal{N}}
\def\cL{\mathcal{L}}

\def\bv{{\bar{v}}}
\def\be{{\bar{e}}}
\def\bp{{\bar{p}}}
\def\sre{\text{SRE}}

\DeclareMathOperator{\Tr}{Tr}

\theoremstyle{plain}
\newtheorem*{theorem*}{Theorem}
\newtheorem{theorem}{Theorem}

\definecolor{tyler}{rgb}{1,.3,0}

\makeatletter
\def\l@subsubsection#1#2{}
\makeatother

\begin{document}

\title{Finite-temperature quantum topological order in three dimensions}

\author{Shu-Tong Zhou}
\affiliation{Department of Physics, Nanjing University,  Nanjing 210093, China}
\author{Meng Cheng}
\affiliation{Department of Physics, Yale University, New Haven, Connecticut  06511-8499, USA}
\author{Tibor Rakovszky}
\affiliation{Department of Physics, Stanford University, Stanford, California 94305, USA}
\affiliation{Department of Theoretical Physics, Institute of Physics,
Budapest University of Technology and Economics, M{\H{u}}egyetem rkp. 3., H-1111 Budapest, Hungary}
\affiliation{HUN-REN-BME Quantum Error Correcting Codes and Non-equilibrium Phases Research Group,
Budapest University of Technology and Economics,
M\H{u}egyetem rkp. 3., H-1111 Budapest, Hungary}
\author{Curt von Keyserlingk}
\affiliation{Department of Physics, King's College London, Strand WC2R 2LS, UK}
\author{{Tyler D.~Ellison}}
\email{tellison@perimeterinstitute.ca}
\affiliation{Department of Physics, Yale University, New Haven, Connecticut  06511-8499, USA}
\affiliation{Perimeter Institute for Theoretical Physics, Waterloo, Ontario N2L 2Y5, Canada}


\begin{abstract}
We identify a three-dimensional system that exhibits long-range entanglement at sufficiently small but nonzero temperature---it therefore constitutes a quantum topological order at finite temperature. The model of interest is known as the fermionic toric code, a variant of the usual 3D toric code, which admits emergent fermionic point-like excitations. The fermionic toric code, importantly, possesses an anomalous 2-form symmetry, associated with the space-like Wilson loops of the fermionic excitations. We argue that it is this symmetry that imbues low-temperature thermal states with a novel topological order and long-range entanglement. Based on the current classification of three-dimensional topological orders, we expect that the low-temperature thermal states of the fermionic toric code belong to a phase of matter that, in the context of equilibrium phases of matter, only exists at nonzero temperatures. We conjecture that further examples of topological orders at nonzero temperatures are given by discrete gauge theories with anomalous 2-form symmetries. Our work therefore opens the door to studying quantum topological order at nonzero temperature in physically realistic dimensions. 
\end{abstract}

\maketitle

In recent decades, one of the central goals
of condensed matter physics has been the classification and characterization of quantum phases of matter at zero temperature. This has led, most notably, to the concept of topological order, where long-range entanglement manifests in exotic properties, such as anyonic excitations. The investigation of zero-temperature topological orders has culminated in conjectured classifications in low dimensions based on sophisticated mathematics~\cite{Kitaev2005anyons, Wen2015theory, Freyd2022classification, Wen2017colloquium, Lan20183Dbosons,Lan20193Dfermions}. 

Despite the theoretical progress, physical experiments cannot be performed in complete isolation from their environment. 
Consequently, systems will inevitably equilibrate to thermal states at nonzero temperature---possibly destroying the zero-temperature order~\footnote{Topological orders in low dimensions can be stabilized far from thermal equilibrium, but this requires highly engineered dynamics (e.g. active error correction). This may be feasible in quantum devices but is less so in conventional solid state systems.}.
A question of fundamental importance is thus: what topological phases of matter can be realized by thermal states above zero temperature? 

It is known that certain topological orders can survive at nonzero temperature; however, to date, all known examples require four spatial dimensions or greater~\cite{Landahl2002TQM, Alicki2010thermalstability4D, Hastings2011nonzero, Bombin2013selfcorrecting, hsin2024nonabelianselfcorrecting}. The prospect of finite-temperature topological order in fewer dimensions has been bleak, with no-go theorems in Refs.~\cite{Nussinov2008fragility, Bravyi2009selfcorrectingnogo, Alicki2009thermalization2d, Yoshida2011selfcorrecting, RobertsSPTnonzeroT2017, Kato2019approximateMarkov, kochanowski2024rapidthermalizationdissipativemanybody, bakshi2025hightemperature, Rouze2024optimalquantumalgorithmgibbs}. This is usually attributed to the fact that these phases possess point-like excitations, which proliferate above zero temperature and destroy the topological order~\cite{Peierls1936Ising, Lan20183Dbosons, Lan20193Dfermions, Castelnovo2008TO3Dtoriccode}. 

Three-dimensional topological orders, such as the 3D toric code, however, exhibit both point-like excitations and loop-like excitations. The latter of which only proliferate at some finite critical temperature~\cite{Peierls1936Ising, Hastings2014Melko}, implying a finite-temperature phase transition. It is tempting to conclude that the topological order persists at nonzero temperature~\cite{Castelnovo2008TO3Dtoriccode}. However, this is not guaranteed to be \textit{quantum} topological order at finite temperature, which is characterized by long-range entangled (LRE) thermal states, according to the definitions in Refs.~\cite{Hastings2011nonzero,Sang2024renormalization}.

Indeed, for the 3D toric code, it was shown in Ref.~\cite{Hastings2011nonzero} that its finite-temperature thermal states are short-range entangled (SRE), i.e., they are classical mixtures of SRE pure states. Therefore, its finite-temperature phase transition is between classical phases~\cite{Lu2020finitetemp}. It was suggested in Ref.~\cite{Hastings2011nonzero} that the same conclusion holds for all finite group gauge theories in 3D. Given that these exhaust a large class of topological orders in 3D~\cite{Lan20183Dbosons,Lan20193Dfermions}, it is important to settle: is it possible to have quantum topological order at nonzero temperature in 3D?

In this letter, we answer this question in the affirmative. We identify a Hamiltonian in 3D with local interactions such that, below a finite-temperature, its thermal states are topologically ordered, 
and moreover cannot be approximated by SRE states.
It therefore constitutes the first example of a finite-temperature topological order in fewer than four spatial dimensions.

Our example is based on the so-called fermionic toric code (\fTC), whose zero-temperature topological order is explored in Refs.~\cite{Levin2003fermion, Keyserlingk2015electrodynamics, Chen2019bosonization3d, Chen2021disentangling}. Similar to the 3D toric code, referred to henceforth as the bosonic toric code (bTC), the {\fTC} is defined on a system of qubits. The key difference from the bTC is that its elementary point-like excitations have fermionic exchange statistics -- i.e., swapping the positions of two of the excitations produces a sign in the pure state wave function. The model furthermore possesses an anomalous 2-form symmetry---simply saying that it is invariant under moving a fermionic excitation along a closed path~\cite{Bhardwaj2017fSPT, Chen2018bosonization, Chen2019bosonization3d, Chen2021disentangling, Ellison2023subsystem}. The symmetry is considered to be `anomalous', because the resulting string operator inherits the fermionic self-statistics. We show that this is 
directly responsible for the long-range entanglement of the low-temperature thermal states. 

We emphasize that, although the {\fTC} can be used as a quantum memory at {zero} temperature, it strictly behaves as a classical memory for nonzero temperatures below a finite-temperature phase transition. This is in contrast to the 4D toric code, which serves as a quantum memory at low temperatures~\cite{Alicki2010thermalstability4D}. 

We begin with a description of the {\fTC} before proving that its low-temperature thermal states are LRE, as made precise in Theorem~\ref{thm: LRE of fTC}.

\begin{figure}[t]
\centering
\includegraphics[width=.48\textwidth]{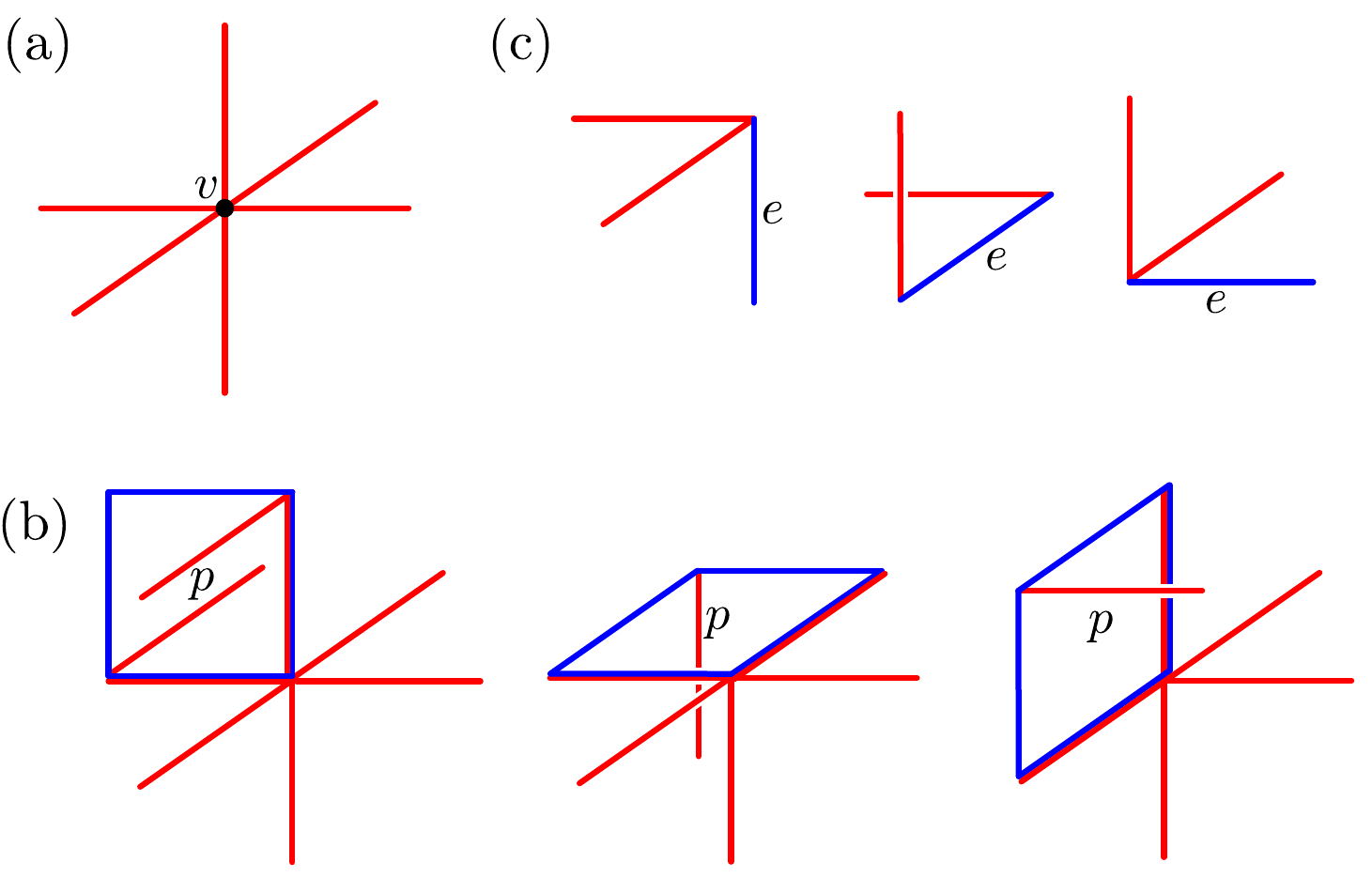}
\caption{Graphical representations of (a) a vertex term $A_v$, (b) plaquette terms $B_p$, and (c) the fermion short string operators $W_e$. Pauli $X$ and $Z$ operators are represented by red and blue edges, respectively. 
}
\label{fig: Hamiltonian terms}
\end{figure}

\textit{Fermionic toric code --}
The {\fTC} is defined on a cubic lattice with a single qubit on each edge, and for simplicity, we assume that the system has periodic boundary conditions. The Hamiltonian can be written as:
\begin{align}
    H = - \sum_v A_v - \sum_p B_p,
\end{align}
where the first sum is over vertices $v$ and the second is over plaquettes $p$. 
We refer to the $A_v$ and $B_p$ as the vertex and plaquette terms, respectively. They are represented graphically in Fig.~\ref{fig: Hamiltonian terms}.

The Hamiltonian terms are mutually commuting and thus define a Pauli stabilizer code~\cite{gottesman1997stabilizer, NielsenChuang2010}. They further satisfy the following relations on a torus:
\begin{align} \label{eq: relations}
    \prod_v A_v = 1, \quad \prod_{p \in c} B_p=1,
\end{align}
where the second product is over the six plaquettes of an arbitrary cube $c$. These relations are identical to those of the bTC, and as such, the models share the same finite-temperature phase transition~\cite{Weinstein2020Xcube}. 

\textit{Excitations--} The {\fTC} exhibits two types of excitations, which correspond to violations of the vertex and plaquette terms. The violations of the $A_v$ terms are point-like and live at the vertices of the lattice. These excitations can be created by operators supported along a path in the lattice, i.e., they are created at the endpoints of string operators. The string operators can be built from the short string operators $W_e$ shown in Fig.~\ref{fig: Hamiltonian terms}.
Notably, in contrast to the bTC, the point-like excitations of the {\fTC} have fermionic exchange statistics, as shown in Ref.~\cite{Chen2019bosonization3d,Chen2021disentangling}.

It is important to note that the Hamiltonian commutes with the string operators defined on closed paths. This defines a 2-form symmetry of the Hamiltonian, i.e., where the symmetry operators are supported on codimension-2 submanifolds. Due to the point-like excitations having fermionic exchange statistics, the 2-form symmetry is anomalous. In other words, there is an obstruction to gauging the symmetry, as this would correspond to condensing the emergent fermions. Note, however, that the symmetry is anomaly free in the presence of physical fermions as these can be bound to the emergent fermions and condensed~\cite{Chen2021disentangling}.

The second type of excitation in the {\fTC} corresponds to violations of the $B_p$ terms. Because of the local relations in Eq.~\eqref{eq: relations}, every cube must have an even number of violated $B_p$ terms. This means that the violated $B_p$ terms form loops in the dual lattice. Accordingly, these excitations are loop-like excitations. 

The loop excitations can be created at the boundaries of membrane operators supported on surfaces in the dual lattice. One choice of membrane operator for a surface $\Sigma$ takes the form $M_\Sigma = \prod_{e \in \Sigma} X_e$.
This operator commutes with all of the $A_v$ terms and only fails to commute with the $B_p$ terms at the boundary of $\Sigma$. 

In analogy to the bTC, the point-like excitations and the loop-like excitations have nontrivial mutual braiding statistics, i.e., moving a fermion around a loop excitation incurs a sign. In other words, string operators that move a fermion along a closed path detect the number of loop excitations enclosed modulo 2. 

\textit{LRE at finite temperature --} We now establish the main result of this work: the low-temperature thermal states of the {\fTC} are LRE, as stated more precisely below:
\begin{theorem} \label{thm: LRE of fTC}
    Below a finite temperature $T_0$, the fidelity \footnote{We take the fidelity of two mixed states $\sigma_1$ and $\sigma_2$ to be: \begin{eqnarray*} 
        F(\sigma_1, \sigma_2) = \Tr\left[ \sqrt{\sqrt{\sigma_1}\sigma_2 \sqrt{\sigma_1}} \right]^2.
    \end{eqnarray*}} between the thermal state $\rho_{\rm{f\hspace{.05em}TC}}$ and an arbitrary SRE state $\sigma_{\rm SRE}$ [see Eq.~\eqref{eq: arbitrary SRE state}] satisfies
    \begin{equation}
        F(\rho_{\rm{f\hspace{.05em}TC}}, \sigma_{\rm SRE})=O(L^{-\infty}),
    \end{equation}
    i.e., the fidelity decays faster than any polynomial in the system size $L$. This implies, via the Fuchs-van de Graaf inequalities, that the trace distance satisfies:
    \begin{align}
       \frac12 \| \rho_{\rm{f\hspace{.05em}TC}} - \sigma_{\rm SRE} \|_1 = 1-O(L^{-\infty}).
    \end{align}
\end{theorem}

We start with an outline for the proof of Theorem~\ref{thm: LRE of fTC}. The temperature $T_0$ is chosen to be low enough that the large loop excitations are suppressed in $\rho_\fTC$. As such, the majority of the loop excitations can be removed with a quasi-local channel (QLC)~\footnote{We take a QLC to be a quantum channel whose dilation is a quasi-local unitary. By a quasi-local unitary, we mean a quantum circuit for which the depth and the size of the supports of the gates are at most polylogarithmic in the system size.}.
This yields a mixed state that well approximates a state with exactly no loop excitations, denoted as $\rho_\varnothing$. 

Next, we leverage the fact that the loop-less state $\rho_\varnothing$ possesses the anomalous 2-form symmetry of the $\fTC$ as a strong symmetry. Meaning that, for any symmetry operator $W$ supported on a contractible path, $\rho_\varnothing$ satisfies $W\rho_\varnothing = \rho_\varnothing$. This is because the symmetry merely detects the parity of loop excitations enclosed by the path, of which there are none for $\rho_\varnothing$. The anomaly then forbids there from being any SRE state capable of approximating $\rho_\varnothing$, as shown recently in Ref.~\cite{Li2024fermionentanglement}. 

Finally, we use that the QLC that cleans up the loop excitations of $\rho_{\fTC}$ maps SRE states to SRE states. If $\rho_{\fTC}$ were well approximated by a SRE state, we would be able to clean up the loop excitations to obtain a SRE approximation to $\rho_\varnothing$ -- a contradiction. In the remainder of this letter, we clarify and fill in the details of the proof. The argument proceeds in three steps.

\textbf{Step one:} The first step is to define a QLC that approximately maps the thermal state $\rho_\fTC$ to the loop-less state $\rho_\varnothing$. 
We accomplish this with a QLC $\cC$, composed of five layers of QLCs. In each layer, we clean up loop excitations on disjoint patches of linear size at least $\ln(L)^2$. After applying all five layers, the only loop excitations that are left behind must have lengths larger than $\ln(L)^2$. These large loop excitations are exponentially suppressed in the thermal state at low temperatures. Thus, the QLC approximately maps $\rho_\fTC$ to $\rho_\varnothing$, in terms of the fidelity of mixed states. We note that, alternatively, a QLC based on the sweep decoder of Ref.~\cite{Kubica2019Sweep} can be used to remove the small loop excitations, as elaborated on in the Supplemental Material (\hyperlink{SM}{SM}). 

\begin{figure*}[t]
\centering
\includegraphics[width=\textwidth]{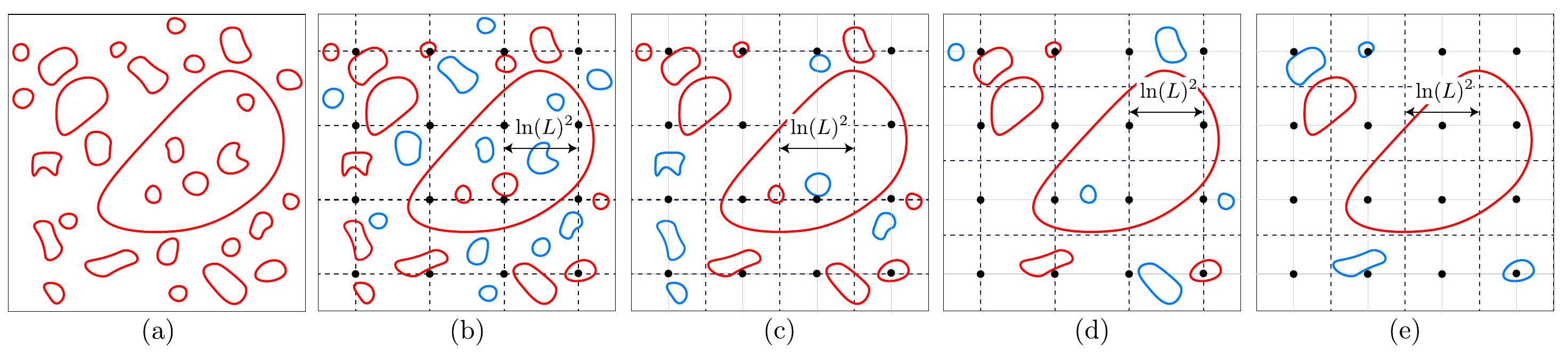}
\caption{2D representation of the QLC $\cC$. (a) A configuration of loop excitations, shown in red. (b) To define the layer $\cC_1$ of the QLC, we divide the system into blocks of linear size at least $\ln(L)^2$ (dashed lines). The loop excitations that are contained entirely within a block are removed with a unitary $U_{\cL_b}$, as in Eq.~\eqref{eq: C Krauss}. These loop excitations are shown in blue. (c) The layer $\cC_2$ is defined by blocking the system (dashed lines) into blocks centered on the vertical seams of the initial blocking (solid lines). The vertices of the initial blocking are marked with a black dot. The loop excitations shown in blue are entirely contained within the second blocking and are removed by $\cC_2$. (d-e) Similarly, the third and fourth layers $\cC_3,\cC_4$ are specified by blocking the system into blocks centered on the horizontal seams or vertices, respectively, of the initial blocking. The remaining loops (red) must have length greater than $\ln(L)^2$.}
\label{fig: loop cleaning}
\end{figure*}

To define $\cC$, we begin by partitioning the system into disjoint blocks of linear size $\ln(L)^2$, shown for a 2D system in Fig.~\ref{fig: loop cleaning}(b). On each of the blocks we apply a quantum channel that removes all loop excitations that are entirely contained within the block. The quantum channel $\cC_b$ on a block $b$ acts as:
\begin{align} \label{eq: C Krauss}
    \cC_b(\cdot) = \sum_{\cL|_b} U_{\cL_b}P_{\cL|_b}\cdot P_{\cL|_b} U_{\cL_b}, 
\end{align}
where $\cL|_b$ is a loop configuration restricted to $b$, $P_{\cL|_b}$ is the projector onto the configuration $\cL|_b$, and $U_{\cL_b}$ is a product of $X$ operators that cleans up all of the loop excitations $\cL_b$ that are fully contained within $b$. The first layer of the QLC $\cC$ is thus $\cC_1 = \bigotimes_b \cC_b$.

Notice that loop excitations that cross between blocks are not removed by $\cC_1$, regardless of their size. Therefore, in the subsequent layers, we clean up loop excitations along the seams of the first blocking, starting with those perpendicular to the $x$-axis, referred to as $x$-seams. We again partition the system into blocks of linear size $\ln(L)^2$, but this time, the blocks are chosen so that they are centered on the $x$-seams of the first blocking, as depicted in Fig.~\ref{fig: loop cleaning}(c). We take the second layer of $\cC$ to be analogous to $\cC_1$ but for the new choice of blocks. We refer to this QLC as $\cC_2$. We repeat this process for the $y$-seams, $z$-seams, and for the corners of the initial blocking. This defines the QLCs $\cC_3$, $\cC_4$, and $\cC_5$.

Altogether, the QLC $\cC$ is ${\cC_5 \circ \cC_4 \circ \cC_3 \circ \cC_2 \circ \cC_1}$. This channel is guaranteed to clean up all of the loop excitations with length less than $\ln(L)^2$. In the \hyperlink{SM}{SM}, we bound the fidelity of $\cC(\rho_\fTC)$ and the state $\rho_\varnothing$ as
\begin{align} \label{eq: fidelity bound cleaned and loopless}
    F(\cC(\rho_{\fTC}), \rho_\varnothing) \geq 1 - L^3e^{-(2 \beta - \ln(5))\ln(L)^2}.
\end{align}
Therefore, if the inverse temperature $\beta$ satisfies $\beta > \frac{\ln(5)}{2}$, the fidelity approaches unity faster than any polynomial in $L$. Below a temperature $T_0 = 2 \ln(5)^{-1}$, we have:
\begin{align} \label{eq: fidelity cleaned and loopless}
    F(\cC(\rho_{\fTC}), \rho_\varnothing) = 1 - O(L^{-\infty}).
\end{align}

We note that the finite-temperature phase transition of the fTC occurs at the critical temperature $T_c \approx 1.31$. This follows from adapting the arguments of Refs.~\cite{Castelnovo2008TO3Dtoriccode,Weinstein2020Xcube} to the fTC, which also imply that the transition belongs to the 3D Ising universality class. The temperature $T_0 \approx 1.24$ is therefore below the critical temperature. However, we expect that the LRE persists up to $T_c$, and that the value of $T_0$ is a consequence of our approximations and a suboptimal algorithm for cleaning up loops. 

\textbf{Step two:} The second step is to argue that $\rho_\varnothing$ cannot be approximated by a SRE state. To make this explicit, we let $\sigma_\sre$ be a mixture of SRE states, such as:
\begin{align} \label{eq: arbitrary SRE state}
    \sigma_\sre = \sum_i p_i |\psi_\sre^i \rangle \langle \psi_\sre^i|.
\end{align}
We assume here that each $|\psi_\sre^i \rangle$ can be prepared from a product state with a local unitary circuit with depth less than $d$, where $d$ is polylogarithmic in the system size. Our goal is to then bound the fidelity between $\rho_\varnothing$ and $\sigma_\sre$ as a function of the depth $d$. 

The fidelity can be immediately bounded using a simple generalization of Corollary 2 in Ref.~\cite{Li2024fermionentanglement} to 3D.  Corollary 2 gives us the bound:
\begin{align}
    F(\rho_\varnothing, \sigma_\sre) < e^{\frac{-cL^3}{(d+1)^3}},
\end{align}
for some positive constant $c$. Note that the only difference from the explicit bounds in Ref.~\cite{Li2024fermionentanglement} are the powers of 3, which are a consequence of 3D. 
If $d$ is polylogarithmic in $L$, we have our desired result:
\begin{align} \label{eq: fidelity loopless and sre}
    F(\rho_\varnothing, \sigma_\sre) = O(L^{-\infty}). 
\end{align}

The absence of loop excitations in $\rho_\varnothing$ means that it obeys a strong 2-form symmetry.
Thus, while the thermal state $\rho_\fTC$ itself is not exactly symmetric, by Eq.~\eqref{eq: fidelity loopless and sre}, the symmetry can be restored (approximately) by cleaning up the loop excitations. Importantly, this symmetry is anomalous, due to the self-statistics of the fermions, which allows us to employ the results of Ref.~\cite{Li2024fermionentanglement}. This is notably distinct from the bTC, whose 2-form symmetry is anomaly free. We again emphasize that in the presence of physical fermions, the 2-form symmetry of the {\fTC} also becomes anomaly free, and hence, $\rho_\varnothing$ can be written as a mixture of SRE states with physical fermions.

\textbf{Step three:} We now upper bound the fidelity between a thermal state $\rho_\fTC$ below the temperature $T_0$ and an arbitrary SRE state $\sigma_\sre$. To begin, we notice that the quantum channel $\cC$ that cleans up small loop excitations can only increase the fidelity, according to the data processing inequality, i.e.,
\begin{align}
    F(\rho_\fTC, \sigma_\sre) \leq F\left(\cC(\rho_\fTC), \cC(\sigma_\sre)\right).
\end{align}

Since $\cC$ is a QLC, the state $\cC(\sigma_\sre)$ is also SRE, up to a tensor product with the identity, as spelled out in the \hyperlink{SM}{SM}.
In particular, there exists a SRE state of the form $\tilde{\sigma}_\sre = \frac{1}{2^{|A|}} \cC(\sigma_\sre) \otimes I_A$, where $|A|$ is the number of ancillary qubits. Taking a tensor product with the identity does not change the fidelity, so we may write:
\begin{align} \label{eq: inequality after cleaning}
    F(\rho_\fTC, \sigma_\sre) \leq F\left(\cC(\rho_\fTC), \tilde{\sigma}_\sre\right).
\end{align}

The bounds in Eqs.~\eqref{eq: fidelity cleaned and loopless} and \eqref{eq: fidelity loopless and sre} suggest that $\cC(\rho_\fTC)$ is close to $\rho_\varnothing$ and that $\rho_\varnothing$ is far from any SRE state.
This then implies that $\cC(\rho_\fTC)$ must be far from any SRE state. This intuition can be made  precise using the triangle inequality for the trace distance and the Fuchs-van de Graaf inequalities~\cite{Fuchs1999inequality}. Combining these two, arbitrary mixed states $\sigma_1$, $\sigma_2$ and $\sigma_3$ must satisfy:
\begin{align} \nonumber
    F(\sigma_2, \sigma_3) \leq
    1- \left[ 1-\sqrt{F(\sigma_1,\sigma_3)}-\sqrt{1-F(\sigma_1,\sigma_2))} \right]^2.
\end{align}

Setting $\sigma_1 = \rho_\varnothing$, $\sigma_2 = \cC(\rho_\fTC)$, and $\sigma_3 = \tilde{\sigma}_\sre$ and 
plugging in the inequalities of Eqs.~\eqref{eq: fidelity cleaned and loopless} and~\eqref{eq: fidelity loopless and sre}, we obtain
\begin{align}
    F(\cC(\rho_\fTC), \tilde{\sigma}_\sre) = O(L^{-\infty}).
\end{align}
Together with Eq.~\eqref{eq: inequality after cleaning}, this gives us our main result and concludes the proof of Theorem~\ref{thm: LRE of fTC}:
\begin{align}
    F(\rho_\fTC, \sigma_\sre) = O(L^{-\infty}).
\end{align}

\textit{Conclusions --} We have demonstrated the existence of finite-temperature quantum topological order in three dimensions, exemplified by the {\fTC} below a temperature ${T_0 = 2 \ln(5)^{-1}}$. The crucial property of the {\fTC}, which enforces long-range entanglement at nonzero temperature, is that it posses an anomalous 2-form symmetry. This is in sharp contrast to the bTC, which has been the focus of previous investigations into finite-temperature topological order in three dimensions~\cite{Castelnovo2008TO3Dtoriccode, Hastings2011nonzero, Lu2020finitetemp}. 

Our argument used that $\rho_\fTC$ at low temperatures can be related to the loop-less state $\rho_\varnothing$ by a QLC. We conjecture that the opposite is also true, and there exists a QLC mapping $\rho_\varnothing$ back to $\rho_\fTC$. This would imply that the two belong to the same mixed-state phase of matter, by the definitions of Refs.~\cite{Sang2024renormalization, Coser2019classification}. Furthermore, due to arguments similar to Refs.~\cite{Wang2025intrinsic,Sohal2025noisy,Ellison2025mixedclassification}, we expect $\rho_\varnothing$ to be inequivalent to any pure state topological order. This would then suggest that the finite-temperature {\fTC} order that we have identified is a quantum phase of matter that exists \emph{only} at nonzero temperature, or in open systems, as a mixed-state topological order~\cite{Fan2024diagnostis, Bao2023errorfielddouble, Wang2025intrinsic, Sohal2025noisy, Ellison2025mixedclassification, Coser2019classification, li2024replica, Sang2024markov, zhang2025strongtoweakspontaneousbreaking1form, ogata2025mixedstate}. 

We expect that further examples of finite-temperature topological orders can be found by generalizing the {\fTC} to other models with anomalous 2-form symmetries. A large class of examples are given by Walker-Wang models based on premodular categories with a transparent fermion~\cite{walker2011WWmodels, Deligne, Ellison2025mixedclassification}.
At an intuitive level, these are discrete $G$ gauge theories obtained by gauging the symmetry of a system with physical fermions. 

The existence of finite-temperature topological order in 3D opens up a number of interesting questions. First, is the long-range entanglement exhibited by the {\fTC} thermal states, and thus the topological order, stable to perturbations of the Hamiltonian? Our arguments used the fact that the {\fTC} Hamiltonian is 2-form symmetric. For a perturbed Hamiltonian, however, the 2-form symmetry is in general emergent, i.e., the system is only guaranteed to have an anomalous 2-form symmetry in its ground-state subspace. Can QLCs be developed to clean up the loop excitations in this case, to (approximately) restore the anomalous 2-form symmetry of the ground state?

Another pressing question is: what diagnostics can be used to detect the topological order of the {\fTC} at nonzero temperature? In principle, one could use a QLC to clean up the loop excitations, then measure the anomalous 2-form symmetry operators in the loop-less state.
Alternatively, it is natural to expect that the long-range entanglement can be detected by computing the topological entanglement negativity, analogous to the 4D toric code in Ref.~\cite{Lu2020finitetemp}. We leave the search for efficiently computable diagnostics of the finite-temperature topological order to future explorations.

Furthermore, it would be interesting to identify physical systems able to realize the finite-temperature topological order of the {\fTC}. The Kitaev model in Ref.~\cite{Mandal2009Kitaev} may be particularly amenable to this pursuit, due to it natively requiring only 2-body interactions. Similar Kitaev-like 2-body interactions have been proposed for neutral atom systems through Floquet engineering in Ref.~\cite{Kalinowski2023Floquet}.  

\vspace{0.2in}
\noindent{\it Acknowledgments} -- TDE is grateful to Timothy H. Hsieh, Shengqi Sang, and Jinmin Yi for valuable conversations about SRE states. Research at Perimeter Institute is supported in part by the Government of Canada through the Department of Innovation, Science and Economic Development and by the Province of Ontario through the Ministry of Colleges and Universities. CvK is supported by UKRI FLF grant MR/Z000297/1.  MC is supported by NSF grant DMR-2424315. TR was supported by Stanford's Q-FARM Bloch Postdoctoral Fellowship in Quantum Science and Engineering and by the HUN-REN Hungarian Research Network through the HUN-REN Welcome Home and Foreign Researcher Recruitment Programme 2023 and through the Supported Research Groups Programme, HUN-REN-BME-BCE Quantum Technology Research Group (TKCS-2024/34).

	
\let\oldaddcontentsline\addcontentsline
\renewcommand{\addcontentsline}[3]{}
\bibliography{bib.bib}
\let\addcontentsline\oldaddcontentsline

\onecolumngrid

\newpage

\appendix

\begin{center}
{\large \bf \hypertarget{SM}{Supplemental Material}} 
\end{center}

\setcounter{equation}{0}
\setcounter{figure}{0}

\setcounter{secnumdepth}{2}

\tableofcontents

\subsection{Sweep decoder}

In the main text, we argued that loop excitations of length $\ln(L)^2$ can be removed with the QLC $\cC$. This channel is comprised of unitaries $U_{\cL_b}$ whose supports have linear size $\ln(L)^2$, as written in Eq.~\eqref{eq: C Krauss}. It may be convenient to instead clean up the loop excitations using a sequence of constant-depth channels. This would only require strictly local operations. However, cleaning up all loops of length $\ln(L)^2$ will necessitate a circuit of depth $O(\ln(L)^2)$. 

Such a strictly local channel is given by the sweep decoder of Ref.~\cite{Kubica2019Sweep}. Broadly speaking, the sweep decoder cleans up loop excitations by incrementally sweeping them in a predetermined direction. We take this direction to be the $(1,1,1)$-direction in the cubic lattice, for simplicity. This is referred to as the causal direction, following Ref.~\cite{Kubica2019Sweep}.

To specify how the loop excitations are swept in the causal direction, it is convenient to work on the dual lattice. The loop excitations then live on the edges of the dual lattice. For a given loop excitation, we define $\mathsf{L}$ to be the set of edges and vertices (in the dual lattice) along the path of the excitation. 
For any vertex $\bv$ in the dual lattice, we further define $\mathsf{L}|_\bv$ to be the edges in $\mathsf{L}$ that are connected to $\bv$, i.e.,
\begin{align}
    \mathsf{L}|_\bv = \{ \be \in \mathsf{L} \,:\, \bv \in \be \}.
\end{align}
We say that a vertex $\bv$ is trailing if it belongs to some loop excitation $\mathsf{L}$ and all of the edges $\be \in \mathsf{L}|_\bv$ are in the positive causal direction from $\bv$, i.e., in the positive $x$, $y$, or $z$ directions from $\bv$. Importantly, any of the vertices of a loop excitation $\mathsf{L}$ that are furthest back in the causal direction are trailing vertices. There may of course be other vertices in $\mathsf{L}$ that are trailing yet are not the vertices that are furthest back in the causal direction.

The way the sweep decoder works is that it identifies the trailing vertices and applies local operators to push those vertices in the causal direction. For example, for the {\fTC}, suppose that the vertex $\bv$ in some loop excitation $\mathsf{L}$ is trailing. The two edges in $\mathsf{L}|_\bv$ share a neighboring plaquette $\bp$ in the dual lattice. Thus, the sweep decoder applies a Pauli $X$ operator on the edge $e$ dual to the plaquette $\bp$. This moves the excitation in the causal direction across the plaquette $\bp$. 

A single round of the sweep decoder can be implemented by a constant-depth channel. Explicitly, for every $B_p$ term, we introduce an ancilla in the $|0\rangle$ state. We then apply local controlled gates to flip the ancilla depending on the eigenvalue of their associated $B_p$ term. In effect, this copies the configurations of the loop excitations onto the ancilla. Next, we apply controlled gates that act with a Pauli $X$ operator on an edge $e$ to move the loop excitation only if the vertex $\bv$ is a trailing vertex. Note that, whether a vertex is trailing or not can be determined locally -- we do not need information about the global loop configuration. Finally, we trace out the ancilla.

A loop excitation of length $\ell$ can be removed with $O(\ell)$ applications of the sweep decoder. To see this, we first note that any loop of length $\ell$ can definitely be contained within a box of dimensions $\ell \times \ell \times \ell$. Given a loop excitation $\mathsf{L}$ of length $\ell$, suppose we have chosen an $\ell \times \ell \times \ell$ box that contains $\mathsf{L}$. We then denote the vertex of the box that is furthest back in the causal direction by $\bv_0$. Every other vertex in the box is in a positive causal direction from $\bv_0$.

After each application of the sweep decoder, the trailing vertices are moved further from $\bv_0$. More precisely, letting $\bar{\mathsf{V}}^{(t)}$ be the set of trailing vertices of the loop excitation after $t$ applications of the sweep decoder, we have:
\begin{align}
  \min_{\bv \in \bar{\mathsf{V}}^{(t+1)}}  \left\{ \text{dist}(\bv_0, \bv) \right\} - \min_{\bv \in \bar{\mathsf{V}}^{(t)}} \{ \text{dist}(\bv_0, \bv)\} \geq 1,
\end{align}
where $\text{dist}(\bv_0, \bv)$ is the taxicab distance between $\bv_0$ and $\bv$.
This tells us that, in particular, the trailing vertices closest to $\bv_0$ are moved away from $\bv_0$ by at least one unit. Furthermore, the taxicab distance between $\bv_0$ and the furthest forward vertex in the box is $3 \ell$. Therefore, it takes at most $3\ell$ applications of the sweep decoder to clean up the loop of length $\ell$. Consequently, loops of length $\ln(L)^2$ can be removed with $3\ln(L)^2$ applications of the sweep decoder, implying that the sweep decoder provides another means of cleaning up loops less than length $\ln(L)^2$ with a QLC.

\subsection{Approximating the loop-less state $\rho_\varnothing$}

To remove the small loop excitations from $\rho_\fTC$, we applied the QLC $\cC$. Here, we show that, at low temperatures, the state $\cC(\rho_\fTC)$ is a good approximation to the loop-less state $\rho_\varnothing$. Specifically, we show that 
\begin{align} \label{eq: fidelity bound cleaned and loopless app}
    F(\cC(\rho_\fTC),\rho_\varnothing) \geq  1 - L^3e^{-(2 \beta - \ln(5))\ln(L)^2},
\end{align}
as claimed in Eq.~\eqref{eq: fidelity bound cleaned and loopless}.

We start by writing $\rho_\fTC$ and $\rho_\varnothing$ explicitly in the {\fTC} eigenbasis. We label the eigenstates as $|\cL, \cV \rangle$, where $\cL$ denotes a configuration of loop excitations, and $\cV$ labels both a configuration of the fermionic excitations and the logical states. In this basis, the thermal state of the {\fTC} is:
\begin{align}
    \rho_\fTC = \frac{1}{\cZ_\cL \cZ_\cV} \sum_{\cL,\cV} e^{-\beta E_\cL} e^{-\beta E_\cV} |\cL, \cV \rangle \langle \cL, \cV|.
\end{align}
Here, the exponents $E_\cL$ and $E_\cV$ denote the energy of the loop excitations and point-like excitations, respectively, in the configuration $\cL$ and $\cV$. We have also introduced the notation $\cZ_\cL$ and $\cZ_\cV$, defined as
\begin{align}
    \cZ_\cL = \sum_\cL e^{-\beta E_\cL}, \quad \cZ_\cV = \sum_\cV e^{-\beta E_\cV}.
\end{align}
The loop-less state is then
\begin{align} \label{eq: varnothing explicit}
   \rho_\varnothing = \frac{1}{\cZ_\cV} \sum_\cV e^{-\beta E_\cV} |\varnothing, \cV \rangle \langle \varnothing, \cV|,
\end{align}
with $\varnothing$ denoting the configuration without any loop excitations. 

We next apply the loop-cleaning channel $\cC$ to $\rho_\fTC$. We let $\cL_\cC$ label the configuration obtained from $\cL$ after acting with $\cC$. The loop-cleaning channel then acts on the eigenstates of the {\fTC} as
\begin{align}
    \cC(|\cL,\cV \rangle \langle \cL, \cV |) = 
    \begin{cases}
        |\varnothing,\cV \rangle \langle \varnothing, \cV |, & \cL \in \mathsf{S}_\varnothing \\
        |\cL_\cC,\cV \rangle \langle \cL_\cC, \cV |, & \cL \not \in \mathsf{S}_\varnothing,
    \end{cases}
\end{align}
where $\mathsf{S}_\varnothing$ denotes the set of loop configurations that are fully cleaned up by $\cC$. Applying $\cC$ to $\rho_\fTC$ produces
\begin{eqs} 
   \cC(\rho_\fTC) = \frac{1}{\cZ_\cL \cZ_\cV} \sum_{\cL \in \mathsf{S}_\varnothing}\sum_{ \cV} e^{-\beta E_\cL}e^{-\beta E_\cV} |\varnothing, \cV \rangle \langle \varnothing, \cV| + \frac{1}{\cZ_\cL \cZ_\cV} \sum_{\cL \not\in \mathsf{S}_\varnothing}\sum_{ \cV} e^{-\beta E_\cL}e^{-\beta E_\cV} |\cL_\cC, \cV \rangle \langle \cL_\cC, \cV |.
\end{eqs}
Notice that the states in the first sum do not depend on $\cL$, therefore we can perform the sum over $\cL \in \mathsf{S}_\varnothing$. This gives us
\begin{eqs} \label{eq: fTC explicit}
   \cC(\rho_\fTC) = \mathbb{P}(\mathsf{S}_\varnothing)\rho_\varnothing + \frac{1}{\cZ_\cL \cZ_\cV} \sum_{\cL \not\in \mathsf{S}_\varnothing}\sum_{ \cV} e^{-\beta E_\cL}e^{-\beta E_\cV} |\cL_\cC, \cV \rangle \langle \cL_\cC, \cV |.
\end{eqs}
Here, $\mathbb{P}(\mathsf{S}_\varnothing)$ is the probability that a loop configuration is cleaned up by $\cC$:
\begin{align}
    \mathbb{P}(\mathsf{S}_\varnothing) = \frac{1}{\cZ_\cL} \sum_{\cL \in \mathsf{S}_\varnothing} e^{-\beta E_\cL}.
\end{align}

We are now prepared to compute the fidelity between $\cC(\rho_\fTC)$ and $\rho_\varnothing$ explicitly. Since both states are diagonal in the {\fTC} eigenbasis, the fidelity simplifies to
\begin{align}
    F(\cC(\rho_\fTC), \rho_\varnothing) 
    = \Tr\left[ \left( \cC(\rho_\fTC) \rho_\varnothing \right)^{\frac12} \right]^2.
\end{align}
Using that the states $|\cL_\cC, \cV \rangle$, for $\cL_\cC \not \in \mathsf{S}_\varnothing$, are orthogonal to states of the form $|\varnothing, \cV \rangle$ in $\rho_\varnothing$,
we obtain
\begin{align}
    F(\cC(\rho_\fTC), \rho_\varnothing) 
    = \Tr\left[\left(\mathbb{P}(\mathsf{S}_\varnothing)\rho_\varnothing^2\right)^{\frac12} \right]^2
    = \mathbb{P}(\mathsf{S}_\varnothing) \Tr[\rho_\varnothing]^2
    = \mathbb{P}(\mathsf{S}_\varnothing).
\end{align}
The probability that a loop configuration can be cleaned up by $\cC$ can alternatively be written as
\begin{align}
    \mathbb{P}(\mathsf{S}_\varnothing) = 1 - \frac{1}{\cZ_\cL} \sum_{\cL \not \in \mathsf{S}_\varnothing} e^{-\beta E_\cL}.
\end{align}
Therefore, the fidelity is
\begin{align} \label{eq: thermal fidelity simplified}
    F(\cC(\rho_\fTC), \rho_\varnothing) = 1 - \frac{1}{\cZ_\cL} \sum_{\cL \not \in \mathsf{S}_\varnothing} e^{-\beta E_\cL}.
\end{align}

To evaluate the bound on the fidelity further, we define $\epsilon$ to be the deviation from unity in Eq.~\eqref{eq: thermal fidelity simplified}:
\begin{align} \label{eq: epsilon definition}
    \epsilon = \frac{1}{\cZ_\cL} \sum_{\cL \not \in \mathsf{S}_\varnothing} e^{-\beta E_\cL}.
\end{align}
We note that the calculation of $\epsilon$ that follows, is similar to Lemma 4 in Ref.~\cite{RobertsSPTnonzeroT2017}.
The first step is to replace the sum over configurations with a sum over energies of loop configurations. This gives us
\begin{align}
    \epsilon = \frac{1}{\cZ_\cL} \sum_{E} \Omega_{\cL \not\in \mathsf{S}_\varnothing}(E) e^{-\beta E},
\end{align}
where $\Omega_{\cL \not\in \mathsf{S}_\varnothing}(E)$ is the multiplicity of configurations $\cL$ with energy $E_\cL = E$ such that, after applying $\cC$, $\cL_\cC$ has at least one loop excitation. 

The multiplicity $\Omega_{\cL \not\in \mathsf{S}_\varnothing}(E)$ is challenging to compute in general. However, we can upper bound it using the fact that all loop excitations of length $\ln(L)^2$ are cleaned up by $\cC$. Therefore, the configurations counted by $\Omega_{\cL \not\in \mathsf{S}_\varnothing}(E)$ must have at least one loop whose length is larger than $\ln(L)^2$. 
This observation allows us to conclude:
\begin{align} \label{eq: bound on multiplicity}
   \Omega_{d} \Omega(E-2d) \geq \Omega_{\cL \not\in \mathsf{S}_\varnothing}(E).
\end{align}
Here, $\Omega_{d}$ is the number of configurations for a single loop of length $d=\ln(L)^2$ and $\Omega(E-2d)$ is the multiplicity of loop configurations of energy $E-2d$. We have subtracted by $2d$, since a loop of length $d$ has energy penalty $2d$.

Plugging the inequality in Eq.~\eqref{eq: bound on multiplicity} into the definition of $\epsilon$, we find:
\begin{eqs}
    \epsilon 
    \leq \frac{1}{\cZ_\cL} \sum_{E \geq 2d} \Omega_{d} \Omega(E-2d) e^{-\beta E}.
\end{eqs}
Note that we have changed the sum to be over energies $E \geq 2d$. This is because $\Omega_{\cL \not\in \mathsf{S}_\varnothing}(E) = 0$ for energies that are less than $2d$. 

Next, we shift the summand by $2d$ and allow the sum to run over all energies of loop configurations. This gives us:
\begin{align}
    \epsilon < \Omega_{d} e^{-\beta 2 d} \frac{1}{\cZ_\cL} \sum_{E} \Omega(E) e^{-\beta E}.
\end{align}
The sum is precisely $\cZ_\cL$, so we find:
\begin{align}
    \epsilon < \Omega_{d} e^{-\beta 2 d}.
\end{align}

The number of configurations $\Omega_{d}$ for a loop of length $d$ is upper-bounded by the number of random walks of length $d$. For a fixed starting point in the cubic lattice, the number of random walks is $5^d$, since at each step, there are 5 possible (non-backtracking) directions to choose. We can further choose any of the $L^3$ sites to start the random walk, so $\Omega_{d}$ is bounded as:
\begin{align}
    \Omega_{d} < L^3 5^{d}.
\end{align}
Therefore, replacing $d$ with $\ln(L)^2$, the bound on $\epsilon$ becomes:
\begin{align}
    \epsilon < L^3 e^{-( 2 \beta - \ln(5)) \ln(L)^2},
\end{align}
which, along with Eq.~\eqref{eq: thermal fidelity simplified}, implies Eq.~\eqref{eq: fidelity bound cleaned and loopless app}.

\subsection{Mapping SRE states to SRE states}

Here, we argue that QLCs map SRE states to SRE states, up to tensoring with the identity on ancillary degrees of freedom \footnote{We thank Shengqi Sang for this observation.}.  To see this, we first note that any QLC can be implemented through the following three operations: (i) tensoring with ancilla in a product state, (ii) applying a quasi-local circuit, and (iii) tracing out a subset of the ancilla. Let $\cC$ be a QLC and $\sigma_\sre$ be an arbitrary SRE state, as in Eq.~\eqref{eq: arbitrary SRE state}. Then, the action of $\cC$ on $\sigma_\sre$ can be expressed as
\begin{align}
    \cC(\sigma_\sre) = \Tr_A\left[ \sum_i p_i V |\psi_\sre^i \rangle \langle \psi_\sre^i| V^\dagger \right],
\end{align}
where a product state has been absorbed into $|\psi_\sre^i\rangle$, $V$ is a quasi-local circuit, and the trace is over the subset of the ancilla $A$. 

Suppose that, prior to tracing out the ancilla, we add depolarizing noise. As a reminder, for a single qubit, the depolarizing noise channel $\cN$ acts on an arbitrary mixed state $\rho$ as 
\begin{align} \nonumber
    \cN(\rho) = \frac14 (\rho + X \rho X + Y \rho Y + Z \rho Z).
\end{align}
We define $\tilde{\sigma}_\sre$ as the state obtained after applying depolarizing noise:
\begin{align}
    \tilde{\sigma}_\sre = \frac{1}{4^{|A|}} \sum_{P_A} \sum_i p_i P_A V |\psi_\sre^i \rangle \langle \psi_\sre^i| V^\dagger P_A.
\end{align}
Here, the sum is over all Pauli operators on the ancilla, and $|A|$ is the number of qubits in $A$. 
As the notation suggests, $\tilde{\sigma}_\sre$ is a SRE mixed state, since each of the pure states $P_A V |\psi_\sre^i \rangle$ is SRE.

Note that, due to the depolarizing noise, $\tilde{\sigma}_\sre$ commutes with every Pauli operator on $A$. Therefore, it must be the identity on $A$ and take the form: $\tilde{\sigma}_\sre = \frac{1}{2^{|A|}} \rho_{\bar{A}} \otimes I_A$, for some mixed state $\rho_{\bar{A}}$ on the complement of $A$. By tracing over $A$, we find:
\begin{align}
    \rho_{\bar{A}} = \Tr_A \left[ \tilde{\sigma}_\sre \right] = \cC(\sigma_\sre).
\end{align}
In the second equality, we used that the depolarizing noise on $A$, prior to tracing out $A$, has no effect on the resulting state. Thus, tensoring $\cC(\sigma_\sre)$ with the identity on $A$ gives $\frac{1}{2^{|A|}}\cC(\sigma_\sre) \otimes I_A = \tilde{\sigma}_\sre$, which is SRE. This shows that, as claimed, QLCs map SRE states to SRE states, up to tensoring with the identity.

\end{document}